\documentclass[twocolumn,secnumarabic,amsmath,amssymb,nobibnotes, aps, prl,superscriptaddress,]{revtex4-2}
\usepackage{graphicx}
\usepackage{dcolumn}
\usepackage{siunitx}
\usepackage{multirow}
\usepackage{color}

\usepackage[mathlines]{lineno}

\begin{document}

\title{Precision mass measurements around ${}^{84}$Mo rule out ZrNb cycle formation \\ in the rapid proton-capture process at type I X-ray bursts}

\author{S.~Kimura}
\email[]{sota.kimura@kek.jp}
\affiliation{Wako Nuclear Science Center, Institute of Particle and Nuclear Studies, High Energy Accelerator Research Organization, Wako 351-0198, Japan}

\author{M.~Wada}
\altaffiliation[Present address: ]{Institute of Modern Physics, Huizhou branch, Chinese Academy of Science, Huizhou 516000, China
}
\affiliation{Wako Nuclear Science Center, Institute of Particle and Nuclear Studies, High Energy Accelerator Research Organization, Wako 351-0198, Japan}

\author{C.Y.~Fu}
\affiliation{RIKEN Nishina Center for Accelerator-Based Science, Wako 351-0198, Japan}
\affiliation{Department of Physics, The University of Hong Kong, Pokfulam 999077, Hong Kong}

\author{N.~Fukuda}
\affiliation{RIKEN Nishina Center for Accelerator-Based Science, Wako 351-0198, Japan}

\author{Y.~Hirayama}
\affiliation{Wako Nuclear Science Center, Institute of Particle and Nuclear Studies, High Energy Accelerator Research Organization, Wako 351-0198, Japan}

\author{D.S.~Hou}
\affiliation{Wako Nuclear Science Center, Institute of Particle and Nuclear Studies, High Energy Accelerator Research Organization, Wako 351-0198, Japan}
\affiliation{Department of Physics, The University of Hong Kong, Pokfulam 999077, Hong Kong}

\author{S.~Iimura}
\affiliation{RIKEN Nishina Center for Accelerator-Based Science, Wako 351-0198, Japan}
\affiliation{Department of Physics, Rikkyo University, Toshima 171-8501, Japan}

\author{H.~Ishiyama}
\affiliation{RIKEN Nishina Center for Accelerator-Based Science, Wako 351-0198, Japan}

\author{Y.~Ito}
\altaffiliation[Present address: ]{Wako Nuclear Science Center, Institute of Particle and Nuclear Studies, High Energy Accelerator Research Organization, Wako 351-0198, Japan
}
\affiliation{Advanced Science Research Center, Japan Atomic Energy Agency, Tokai, 319-1195, Japan.}

\author{S.~Kubono}
\affiliation{RIKEN Nishina Center for Accelerator-Based Science, Wako 351-0198, Japan}
\affiliation{Center for Nuclear Study, The University of Tokyo, Wako, 351-0198, Japan}

\author{K.~Kusaka}
\affiliation{RIKEN Nishina Center for Accelerator-Based Science, Wako 351-0198, Japan}

\author{S.~Michimasa}
\affiliation{RIKEN Nishina Center for Accelerator-Based Science, Wako 351-0198, Japan}

\author{H.~Miyatake}
\affiliation{Wako Nuclear Science Center, Institute of Particle and Nuclear Studies, High Energy Accelerator Research Organization, Wako 351-0198, Japan}

\author{S.~Nishimura}
\affiliation{RIKEN Nishina Center for Accelerator-Based Science, Wako 351-0198, Japan}

\author{T.~Niwase}
\affiliation{Department of Physics, Kyushu University, Fukuoka, 819-0395, Japan}

\author{V.~Phong}
\affiliation{RIKEN Nishina Center for Accelerator-Based Science, Wako 351-0198, Japan}

\author{M.~Rosenbusch}
\affiliation{RIKEN Nishina Center for Accelerator-Based Science, Wako 351-0198, Japan}

\author{H.~Schatz}
\affiliation{Facility for Rare Isotope Beams and Department of Physics and Astronomy, Michigan State University, East Lansing, Michigan 48824 USA}

\author{P.~Schury}
\affiliation{Wako Nuclear Science Center, Institute of Particle and Nuclear Studies, High Energy Accelerator Research Organization, Wako 351-0198, Japan}

\author{Y.~Shimizu}
\affiliation{RIKEN Nishina Center for Accelerator-Based Science, Wako 351-0198, Japan}

\author{H.~Suzuki}
\affiliation{RIKEN Nishina Center for Accelerator-Based Science, Wako 351-0198, Japan}

\author{A.~Takamine}
\affiliation{RIKEN Nishina Center for Accelerator-Based Science, Wako 351-0198, Japan}
\affiliation{Department of Physics, Kyushu University, Fukuoka, 819-0395, Japan}

\author{H.~Takeda}
\affiliation{RIKEN Nishina Center for Accelerator-Based Science, Wako 351-0198, Japan}

\author{Y.~Togano}
\affiliation{RIKEN Nishina Center for Accelerator-Based Science, Wako 351-0198, Japan}

\author{Y.X.~Watanabe}
\affiliation{Wako Nuclear Science Center, Institute of Particle and Nuclear Studies, High Energy Accelerator Research Organization, Wako 351-0198, Japan}

\author{W.D.~Xian}
\affiliation{Wako Nuclear Science Center, Institute of Particle and Nuclear Studies, High Energy Accelerator Research Organization, Wako 351-0198, Japan}
\affiliation{Department of Physics, The University of Hong Kong, Pokfulam 999077, Hong Kong}

\author{Y.~Yanagisawa}
\affiliation{RIKEN Nishina Center for Accelerator-Based Science, Wako 351-0198, Japan}

\author{T.T.~Yeung}
\affiliation{RIKEN Nishina Center for Accelerator-Based Science, Wako 351-0198, Japan}
\affiliation{Department of Physics, The University of Tokyo, Bunkyo 113-0033, Japan}

\author{M.~Yoshimoto}
\affiliation{RIKEN Nishina Center for Accelerator-Based Science, Wako 351-0198, Japan}

\author{S.~Zha}
\affiliation{RIKEN Nishina Center for Accelerator-Based Science, Wako 351-0198, Japan}
\affiliation{Department of Physics, The University of Hong Kong, Pokfulam 999077, Hong Kong}

\date{\today}

\begin{abstract}
The rapid proton-capture ($rp$-) process is one of the primary, explosive thermonuclear burning processes that drive type I X-ray bursts. A possible termination of the $rp$-process at around ${}^{84}$Mo was previously suggested by the formation of a ZrNb cycle. We report here precision mass measurements at around ${}^{84}$Mo, which have concluded the possibility of the cycle. The experiment was conducted using the multi-reflection time-of-flight spectrograph at RIKEN RI Beam Factory, and the masses of  ${}^{79}$Y, ${}^{83}$Nb, ${}^{84}$Mo, ${}^{88}$Ru, and an isomer in ${}^{78}$Y were measured. For ${}^{84}$Mo, and ${}^{88}$Ru, and the isomeric state of ${}^{78}$Y, their masses are experimentally determined for the first time with uncertainties of  $\delta m \approx 20~{\rm keV/c^2}$.  The mass precision of ${}^{79}$Y and ${}^{83}$Nb is improved to $13~{\rm keV/c^2}$ and $9.6~{\rm keV/c^2}$, respectively. The new $\alpha$-separation energy of ${}^{84}$Mo, 1.434(83)~MeV, unambiguously rules out the possibility of forming the ZrNb cycle. The X-ray burst simulation with the new masses shows that our measurements effectively remove the large final abundance uncertainties in the $A=80-90$ mass region. The new mass values improve the prediction power for the composition of the nuclear ashes in X-ray bursts.
\end{abstract}

\maketitle


Type I X-ray bursts are known as the most frequent explosive astronomical phenomenon in the universe, occurring on the surfaces of neutron stars in low-mass X-ray binary systems \citep{Parikh2013, Meisel2018, Galloway2021}. X-ray bursts are explained by unstable thermonuclear burning of matter accreted from a stellar companion. The dominant reaction sequence in common with mixed hydrogen and helium bursts is a rapid proton-capture process ($rp$-process), which will proceed away from the valley of stability and reach the proton-drip line in the intermediate mass region on the nuclear chart. In the case that the accreted material composition is hydrogen-rich and low-metallicity, the $rp$-process will proceed up to the vicinity of the ${}^{100}$Sn \citep{Schatz2001, Jose2010}. 

The observables of X-ray bursts are light curves and recurrence times; comparisons between observations and theoretical predictions can be used to extract information about the neutron star mass and radius \citep{Zamfir2012, Meisel2019}, constraining the nuclear matter equation of state \citep{Ozel2016, Xie2024}. X-ray burst models that match observations are also needed to predict the composition of the nuclear burst ashes. This is important for the prediction of any observable spectral signatures of nucleosynthesis contributions from ejected material \citep{Weinberg2006, Herrera2023} and of the composition of the accreted crust. The latter is critical for the interpretation of observations of crustal cooling in quasi-persistent soft X-ray transients \citep{Meisel2018}. However, nuclear physics uncertainties, as they are addressed in this work, significantly limit current X-ray burst models and, thus, the interpretation of observations and predictions of observables \citep{Parikh2008, Cyburt2016}.

One of the nuclear physics effects with a particularly strong impact on the $rp$-process is closed reaction cycles, which will regulate or stop the $rp$-process reaction flow. In the region above the ${}^{56}$Ni on the nuclear chart, several cycles, the NiCu, the ZnGa, the GeAs cycles \citep{Wormer1994}, the ZrNb cycle \citep{Schatz1998}, and the SnSbTe cycle \citep{Schatz2001}, have been predicted. These reaction cycles will be established by enhancing the $\alpha$-evaporation channel in the proton-capture reaction instead of the deexcitation with $\gamma$-ray emission or by strong $(\gamma,\alpha)$ photodisintegration at high temperatures. This enhancement of the $\alpha$-channel is due to the small $\alpha$-separation energies ($S_{\alpha}$) of ${}^{60}$Zn, ${}^{64}$Ge, ${}^{68}$Se, ${}^{84}$Mo, and ${}^{107,108}$Te. For ${}^{84}$Mo, the prediction is based on theoretical masses. Thus, accurate mass data of these key nuclides are needed to clarify the existence and the strength of these cycles in the $rp$-process for given environmental conditions. Mass measurement studies focusing on these cycles were previously reported in \citep{Block2006, Elomaa2009, Kankainen2010, Haettner2011, Xing2018, Xing2023, Ireland2025}.

\begin{figure}[t]
\centering 
\begin{center}
\includegraphics[width=0.4\textwidth,  bb = 0 0 842 595, clip, trim=70 20 90 0]{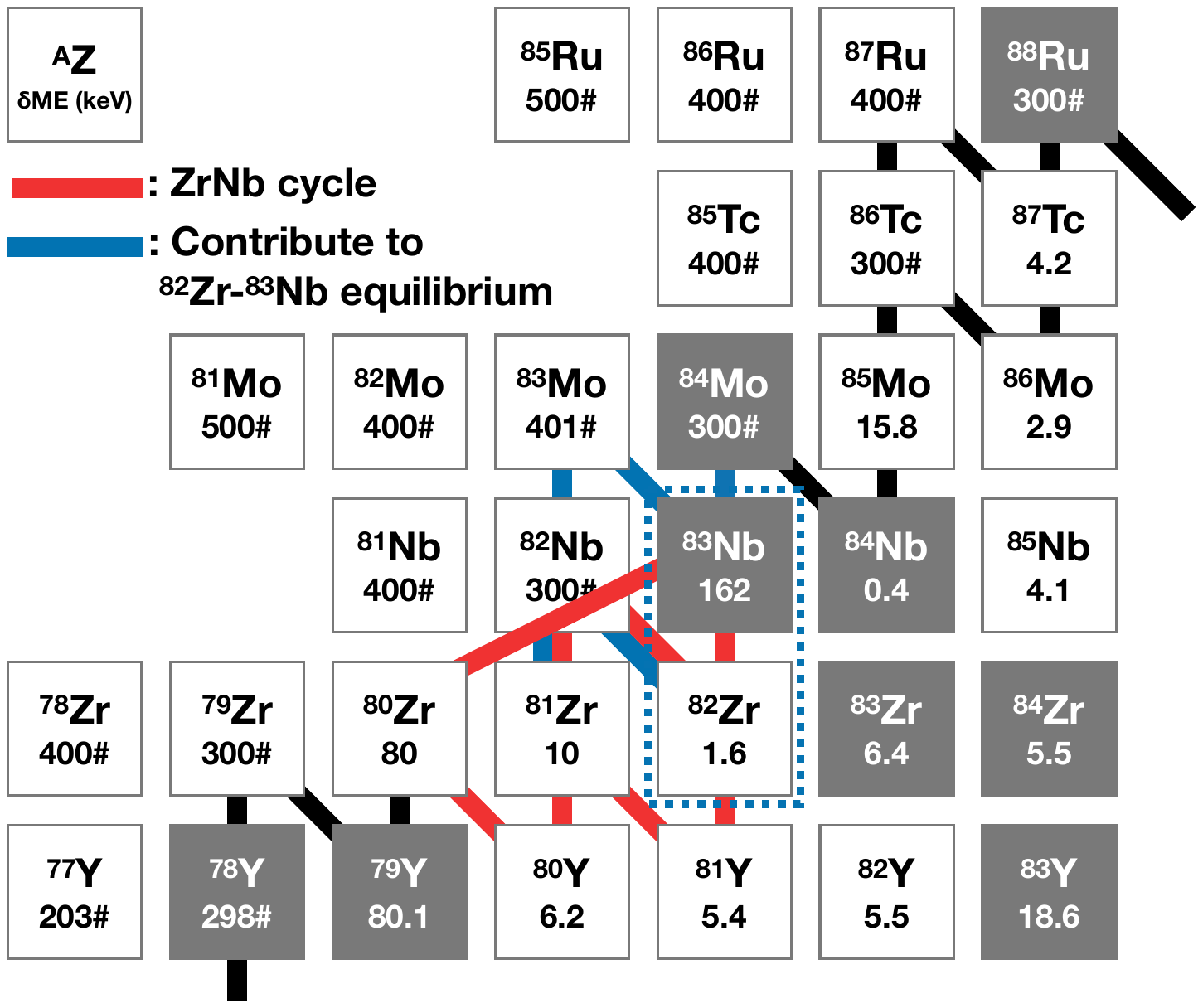}
\end{center}
\caption{Nuclear chart in the region of interest. The uncertainty of the mass excess is presented below the symbol of a nuclide; the \# symbol is given for the AME20's extrapolation value. The nuclides observed in the present measurements are indicated with the grey-colored boxes. The thick lines represent the predicted $rp$-process path. The blue dotted-line box highlights the ${}^{82}{\rm Zr}$ - ${}^{83}{\rm Nb}$ pair, which significantly affects the final abundance. \label{rpPath}}
\end{figure}

\begin{figure*}[t]
\centering 
\begin{center}
\includegraphics[width=1.\textwidth,  bb = 0 0 362 172, clip, trim=0 0 0 0]{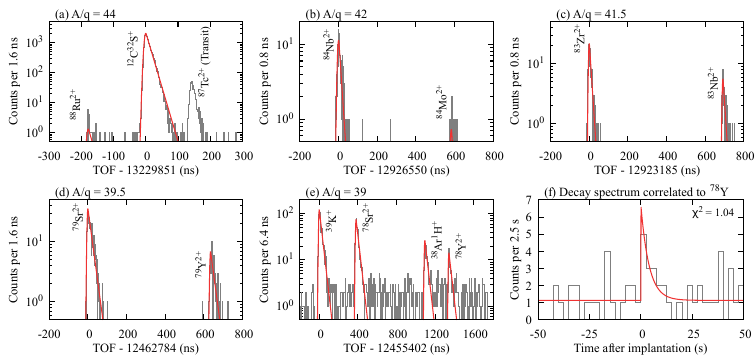}
\end{center}
\caption{(a) -- (e): The observed time-of-flight spectra. The red-colored lines indicate the fit results. The enlarged, partial spectra are presented for (b) and (c). (f): $\beta$-decay time spectrum correlated to the ${}^{78}{\rm Y}$'s TOF events. The red line represents the fit result with a decay curve + constant background.\label{TOFspec}}
\end{figure*}

The ZrNb cycle would be established by a dominant $\alpha$-emission channel in the proton capture reaction of ${}^{83}{\rm Nb}$ (Fig.~\ref{rpPath}). It would strongly limit the synthesis of elements beyond $A \sim 84$ in the $rp$-process, preventing, for example, any significant contribution to the light $p$-nuclei in the $A=92-98$ mass range. The possible existence of the ZrNb cycle has been postulated based on the very low ${}^{84}{\rm Mo}$'s $S_{\alpha}$ of $-0.58~{\rm MeV}$ predicted by the finite range liquid drop model, FRDM92 \citep{Moller1995}, that would make ${}^{84}{\rm Mo}$ $\alpha$-unbound \citep{Schatz1998}. While subsequent updates of the FRDM mass model predict substantially larger values, the existence and formation temperature of the cycle depends on the exact masses of ${}^{80}{\rm Zr}$, ${}^{83}{\rm Nb}$, and ${}^{84}{\rm Mo}$. For ${}^{83}{\rm Nb}$, the mass was measured via isochronous mass spectrometry in the CSRe of Lanzhou, and its mass excess (ME) was determined to be ${\rm ME} = - 57613(162)~{\rm keV}$ \cite{Xing2018}. This value is $\approx 800~{\rm keV}$ less bound than the extrapolation of the 2012 Atomic Mass Evaluation \citep{Wang2012}. The authors have concluded that the formation of the ZrNb cycle is unlikely with the extrapolated mass values based on their new masses of ${}^{81}{\rm Zr}$ and ${}^{83}{\rm Nb}$. However, the measurement of the mass of ${}^{80}{\rm Zr}$ with Penning trap mass spectrometry at the LEBIT facility at MSU shows that its ground state is unexpectedly more bound than predicted \cite{Hamaker2021}. This leads to the possibility that ${}^{84}{\rm Mo}$ has a lower than expected $S_{\alpha}$. The prediction of the masses of $N=Z$ nuclei is particularly challenging owing to additional pairing effects  \citep{Wang2023, Wang2024}. An experimental determination of the ${}^{84}{\rm Mo}$ mass is the last piece of information needed to conclusively determine whether the ZrNb cycle in the $rp$-process forms at X-ray burst conditions. 

Typically, mass accuracies of better than $10~{\rm keV/c^2}$ are needed to not affect burst models significantly \citep{Schatz2006, Schatz2013}. In the vicinity of ${}^{84}$Mo, there are additional mass uncertainties in the $N=42$ isotones that prevent the determination of additional impedances of the $rp$-process reaction flow beyond the $A=40-84$ mass region. The proton-separation energy ($S_p$) of ${}^{83}$Nb ranges $S_p = 802 - 1774~{\rm keV}$ within $\pm 3\sigma$ deviation. The lower-side value allows ${}^{82}{\rm Zr}$ to behave like a waiting point, and the larger side is not. The points in the $rp$-process path where the local $(p,\gamma)$-$(\gamma,p)$ equilibrium is established are called waiting points, because the reaction flow ``waits" at this point until a slower, alternative reaction such as $\beta^+$-decay allows further progress. Thus, the reaction flow strongly depends on the presence of the waiting-point nuclei, and thus on $S_p({}^{83}{\rm Nb})$. 

\begin{table*}[t]
\caption{Measured square of time-of-flight ratio $\rho^2$ and the extracted mass excess values ME. The nuclides used as atomic mass references are shown in column ``Reference". ME$_{\rm lit}$ indicates the literature's mass excess values, and $\Delta {\rm ME}$ represents the differences between ME$_{\rm lit}$ and the mass excess values of the present study: $\Delta {\rm ME} \equiv {\rm ME} - {\rm ME}_{\rm lit}$. The extrapolation value of AME20 is indicated with the \# symbol. \label{Summary}}
\renewcommand{\arraystretch}{1.5}
\begin{ruledtabular}
\begin{tabular}{cclcSSS}
\textrm{Species}&
\textrm{Reference}&
\multicolumn{1}{c}{\textrm{$\rho^2$}}&
\multicolumn{1}{c}{\textrm{$\delta (\rho^2) / \rho^2$}}&
\multicolumn{1}{c}{\textrm{ME~(keV)}}&
\multicolumn{1}{c}{\textrm{ME$_{\rm lit.}$~(keV)}}&
\multicolumn{1}{c}{\textrm{$\Delta {\rm ME}$~(keV)}} 
\\ \hline
\colrule
${}^{88}{\rm Ru}^{2+}$ & ${}^{12}{\rm C}{}^{32}{\rm S}^{+}$ &0.99997291(23)&$2.3 \times 10^{-7}$&-54250(19)&-54340(300)\hspace{-1.5em}\#&90(301)\# \\
${}^{84}{\rm Mo}^{2+}$ & ${}^{84}{\rm Nb}^{2+}$ &1.00009026(28)&$2.8 \times 10^{-7}$&-54137(22)&-54170(298)\hspace{-1.5em}\#&33(299)\# \\
${}^{84}{\rm Zr}^{2+}$ & ${}^{84}{\rm Nb}^{2+}$ &0.99986897(37)&$3.7 \times 10^{-7}$&-71438(29)&-71421.7(55)&-16(30)\\
${}^{83}{\rm Nb}^{2+}$ & ${}^{83}{\rm Zr}^{2+}$ &1.000107220(92)&$9.2 \times 10^{-8}$&-57629.2(96)&-57613(162)&-16(162)\\
${}^{83}{\rm Y}^{{\rm m}~2+}$ & ${}^{83}{\rm Zr}^{2+}$ &0.99991944(16)&$1.6 \times 10^{-7}$&-72135(14)&-72144(18)&9(23)\\
${}^{79}{\rm Y}^{2+}$ & ${}^{79}{\rm Sr}^{2+}$ &1.00010196(14)&$1.4 \times 10^{-7}$&-57984(13)&-57803(80)&-181(81)\\
${}^{78}{\rm Y}^{{\rm m}~2+}$ & ${}^{39}{\rm K}^{+}$ &1.00021567(20)&$2.0 \times 10^{-7}$&-51959(15)& & \\
 &  & \multicolumn{3}{r}{\textrm{Comparison with the ground state:}} & -52173(298)\hspace{-1.5em}\#& 214(298)\#\\
${}^{78}{\rm Sr}^{2+}$ & ${}^{39}{\rm K}^{+}$ &1.00006126(12)&$1.2 \times 10^{-7}$&-63168.0(89)&-63174.0(75)&6.0(116)\\
${}^{38}{\rm Ar}{}^{1}{\rm H}^{+}$ & ${}^{39}{\rm K}^{+}$ &1.00017569(19)&$1.9 \times 10^{-7}$&-27430.7(70)&-27425.86(20)&-4.9(70)\\
\end{tabular}
\end{ruledtabular}
\renewcommand{\arraystretch}{1.0}
\end{table*}


In this letter, we present the results of mass measurements of the proton-rich nuclei in the vicinity of ${}^{84}$Mo with a multi-reflection time-of-flight mass spectrograph (MRTOF-MS) \citep{Schury2014}. The experiment was performed using the RIKEN-KEK Collaborative RI Stopper and Mrtof-based Analyzer and Spectroscopy System (CRISMASS), which is composed of a radio-frequency carpet-type helium gas catcher (RFGC) \citep{Takamine2019} and the MRTOF-MS (ZD-MRTOF) \citep{Rosenbusch2023} in the SLOWRI project \citep{Iimura2023}. It is located downstream of the ZeroDegree Spectrometer (ZDS) at BigRIPS \citep{Kubo2003, Kubo2012} in RIBF \citep{Yano2007}. The measured proton-rich unstable nuclei were produced via the fragmentation reaction of a 345~MeV/nucleon ${}^{124}$Xe beam. They were stopped and converted to extremely low-energy ions using the RFGC behind the ZDS. The ions extracted from the RFGC were transported to a planar-geometry radiofrequency quadrupole trap \citep{Ito2013b}, an injector for the MRTOF-MS. After that, their time-of-flight (TOF) spectra were measured with the MRTOF-MS. A $\beta$-TOF detector \citep{Niwase2023}, which enables the simultaneous measurement of the TOF signals and the $\beta$-decay events, was used as an ion detector of the MRTOF-MS.

The analysis process was the same as described in \citep{Kimura2024}, and the single reference method \citep{Ito2013} was employed for extracting mass values; the ion mass $m_{\rm X}$ is given by $m_{\rm X} = \rho^2 m_{\rm ref}$, where $m_{\rm ref}$ is the mass of the reference ion and $\rho$ is the TOF ratio defined by $\rho \equiv (t_{\rm X} -t_0) / (t_{\rm ref} -t_0)$. $t_{\rm X}$ and $t_{\rm ref}$ are the TOF of the analyte and the mass reference, respectively, and $t_0$ is the systematic offset depending on the measurement system. In the present measurements, it was determined to be $t_{0} = 220(30)~{\rm ns}$ form the TOF ratio of ${}^{84}{\rm Nb}^{2+}$/${}^{39}{\rm K}^{+} $ pair by using the concomitant referencing method \citep{Schury2018}. The isobaric mass references  with the same $A/q$ were used for mass determination. With this, the uncertainty caused by the error of $t_0$ is $ \delta (\rho^2)_{\rm sys} /\rho^2 \lesssim 10^{-9}$ \citep{Kimura2021} and is negligible in the present study. In addition, another uncertainty originating from the binning process, $ \delta \rho_{\rm sys, bin} = 1 \times 10^{-8}$ \citep{Kimura2018}, was accounted for in the error evaluation.


\begin{figure}[t]
\centering 
\begin{center}
\includegraphics[width=0.5\textwidth,  bb = 0 0 362 362, clip, trim=0 0 0 0]{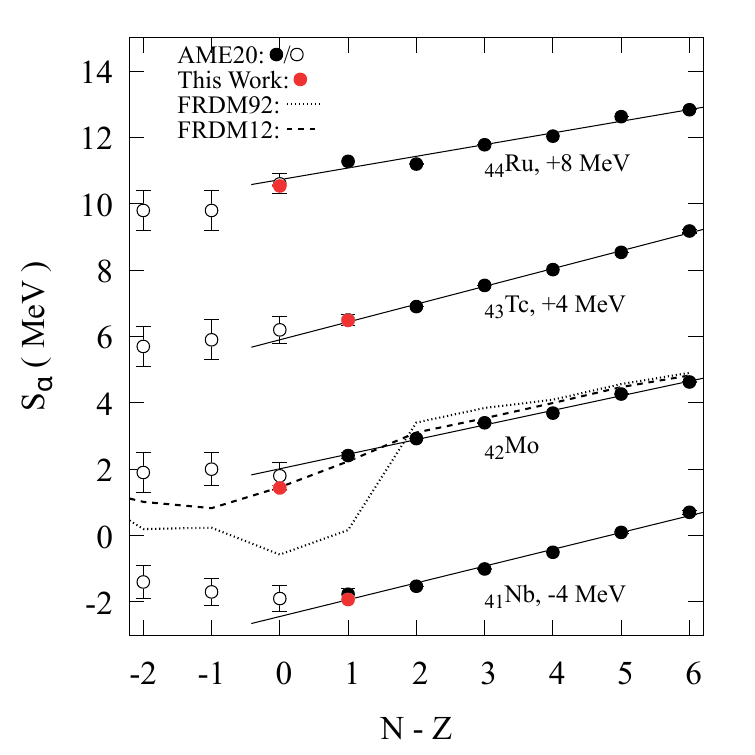}
\end{center}
\caption{$\alpha$-separation energy of the Nb, Mo, Tc, and Ru isotope series. The filled circles indicate the experimental values evaluated in AME20 and based on the present measurements, while the open symbols represent the extrapolated values. The linear fit results with the known AME20 data are given as guides to the eye. For Mo, theoretical predictions of FRDM92 and FRDM12 are also plotted. \label{aSeparation}}
\end{figure}

\begin{figure}[t]
\centering 
\begin{center}
\includegraphics[width=0.5\textwidth,  bb = 0 0 1088 1088, clip, trim=0 0 0 0]{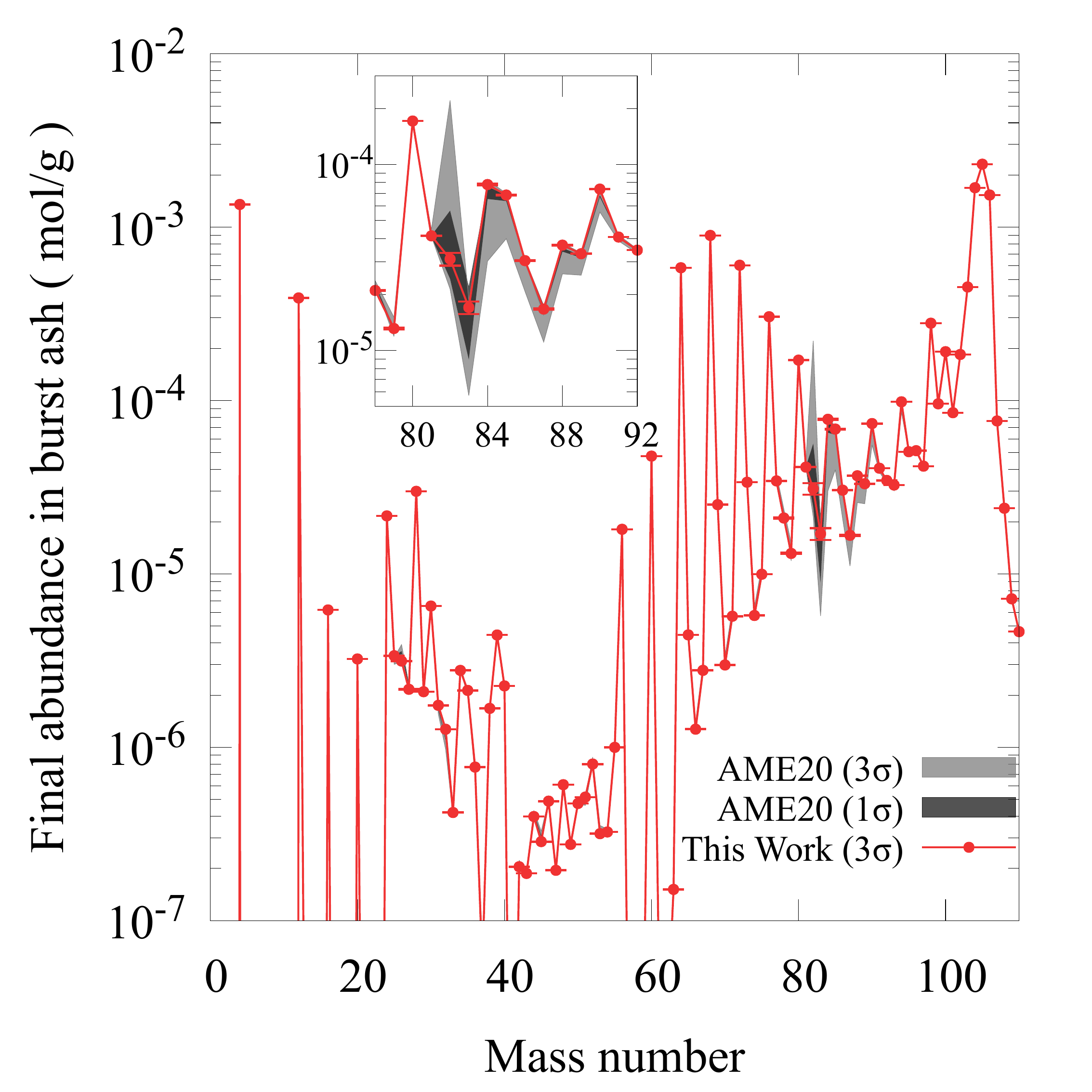}
\end{center}
\caption{Abundance plot of the simulated X-ray burst ash. The grey and dark-grey bands represent the range of variation of the abundance pattern by changing the masses of the four nuclides within $3\sigma$ and $1\sigma$ of the AME20 data. \label{abundance}}
\end{figure}

The measurements were done for the doubly charged $A = 88$, 84, 83, 79, and 78 isobar series. The observed TOF spectra are shown in Fig.~\ref{TOFspec} (a)-(e), and the fifteen ion species, including the molecules of the stable isotopes, were identified. The peak of ${}^{87}{\rm Tc}^{2+}$ was identified as a transient contaminant in Fig.~\ref{TOFspec} (a). No other ion with $A/q=43.5$ was observed, and the single reference method cannot be applied; therefore, ${}^{87}{\rm Tc}$ is excluded from the discussions in this letter.

The peak of ${}^{78}$Y$^{2+}$ was found in the $A/q= 39$ series (Fig.~\ref{TOFspec} (e)). The two $\beta$-decaying states have been known for ${}^{78}$Y; one is the ground state with the half-life ($T_{1/2}$) of $T_{1/2} = 54(5)~{\rm ms}$, and another is an isomeric state of $T_{1/2} = 5.8(6)~{\rm s}$, whose the excitation energy is unknown. The $\beta$-decay information was used to confirm the state of the observed ${}^{78}$Y. To get the TOF-correlated $\beta$-decay events, the TOF events were selected by the fit process with the TOF spectrum first, and then all $\beta$-decay events that occurred within the time window centered on the selected TOF events were collected. A fit function consisting of a decay curve, including the contribution from the daughter nuclide, and the constant background was employed for half-life determination. Figure~\ref{TOFspec} (f) shows the correlated $\beta$-decay events, and the measured half-life is $2.8^{+2.4}_{-1.3}~{\rm s}$. This is consistent with the isomeric state. Then, we have concluded that the observed peak corresponds to the isomeric state of ${}^{78}$Y. 

The analysis results are summarized in Table~\ref{Summary}. All reference mass values were adopted from the 2020 Atomic Mass Evaluation (AME20) \citep{Huang2021, Wang2021}. The mass resolving power of the present measurement was $R_{\rm m} \approx 6.5 \times 10^5$, and the masses of the proton-rich, $N \geq Z$ nuclides are experimentally determined with uncertainties of $\delta m \approx 10-20~{\rm keV/c^2}$. The masses of ${}^{88}$Ru, ${}^{84}$Mo, and ${}^{78}$Y${}^{\rm m}$ are determined for the first time. For ${}^{83}$Nb, the uncertainty of its mass excess value is reduced to 9.6~keV from the literature value of 162~keV: the mass precision is improved by a factor of 17. The disagreement with the literature value is found for ${}^{79}$Y. The present adopted value is taken from the mass measurement study with the CSRe \cite{Xing2018} and is 181~keV heavier than the present result. Due to its shorter measurement time of $200~\mu{\rm s}$ \cite{Xing2018}, this difference could be accounted for if an unknown, short-lived isomeric state in ${}^{79}$Y exists. Thus, we propose the new mass excess value of ${}^{79}$Y: ${\rm ME} = -57984(13)~{\rm keV}$. As discussed above, the excitation energy ($E_{\rm X}$) of the ${}^{78}$Y's isomeric state has been unknown yet, but the $E_{\rm X} \lesssim 500~{\rm keV}$ is suggested from the decay spectroscopy study \citep{Uusitalo1998}. If this estimation is reasonable, the ground state of ${}^{78}$Y is approximately 300 keV more bound than expected. The results for the other ion species are consistent with the literature values within $1\sigma$. 

Figure~\ref{aSeparation} plots the $S_{\alpha}$ of the Nb, Mo, Tc, and Ru isotope series. A sudden drop of $S_{\alpha}$ at ${}^{84}$Mo from the linear trend is observed clearly in the Mo case, in contrast with the other three isotope series showing good linearity in the experimentally known region, including the present measurements. This is the first evidence of the existence of the pronounced island of low $S_{\alpha}$ in the neutron-deficient Mo isotopes, which was initially claimed to start at ${}^{85}$Mo \cite{Haettner2011} while its existence was later rejected \cite{Xing2018}.

The present study determines the $S_{\alpha}$ of ${}^{84}$Mo for the first time with experimental data, resulting in $S_{\alpha} =1.434(83)~{\rm MeV}$. The FRDM12  \citep{Moller2016} reproduces the trend of Mo's $S_{\alpha}$ well despite the difference in the experimental and the theoretical mass excess values of Zr and Mo isotopes, which reaches around 0.6~MeV at $N=Z$. The prediction of the FRDM92, which led to predictions of the existence of the ZrNb cycle \citep{Schatz1998}, can be excluded according to the present $S_{\alpha}({}^{84}{\rm Mo})$ value.  With the present value, we can conclude that the Zr-Nb cycle does not form in the realistic $rp$-process scenarios in X-ray bursts. 

An X-ray burst simulation was carried out to confirm the impact of the new mass values on the $rp$-process. We used a one-zone model that simulates nuclear processes in an extremely hydrogen-rich X-ray burst with a particularly extended rp-process. The model is described in more detail in \citep{Schatz2001}. It uses an accreted composition characterized by solar proportions of hydrogen and hellium and a metallicity of 10$^{-3}$, and an accretion rate of $8.8 \times 10^3~{\rm g/ (cm^2 \cdot s)}$. The calculations were performed by varying the mass values of ${}^{79}{\rm Y}$, ${}^{83}{\rm Nb}$, ${}^{84}{\rm Mo}$, and ${}^{88}{\rm Ru}$ up and down within $3\sigma$ in all combinations. The proton capture rate and its reverse rates were recalculated using TALYS \citep{Koning2023} for each calculation according to the change in mass value.

While we do not find any impact on the burst light curve because the present focused region corresponds to the late stage of the bursts (near the end of light curve), the large AME20 mass uncertainties prior to our experiment lead to significant uncertainties in the final abundances (Fig.~\ref{abundance}). At the 3$\sigma$ level, these reach factors of 10 and 5.5 for $A=82$ and $A=83$, respectively. The reason is that the AME20 masses allow a proton separation energy of ${}^{83}{\rm Nb}$ as low as $800~{\rm keV}$. This leads to the establishment of ($p,\gamma$)-($\gamma,p$) equilibrium between ${}^{82}{\rm Zr}$ and ${}^{83}{\rm Nb}$, effectively making ${}^{82}{\rm Zr}$ an $rp$-process waiting point. Unlike other $rp$-process waiting points, it is fed via ${}^{81}{\rm Zr}(p,\gamma){}^{82}{\rm Nb}(\beta^+){}^{82}{\rm Zr}$ and ${}^{81}{\rm Zr}(2p,2\gamma){}^{83}{\rm Mo}(\beta^+){}^{83}{\rm Nb}$ (Fig.~\ref{rpPath}). The ${}^{84}{\rm Mo}$ mass uncertainty also contributes. The AME20 mass uncertainties allow for a significantly reduced $S_p({}^{84}{\rm Mo})$, which leads to a reduced ${}^{83}{\rm Nb}(p,\gamma){}^{84}{\rm Mo}$ reaction rate, suppressing leakage out from the ($p,\gamma$)-($\gamma,p$) equilibrium of ${}^{82}{\rm Zr}$-${}^{83}{\rm Nb}$. Then, the AME20 ${}^{84}{\rm Mo}$ mass uncertainty therefore also contributes to the $A=82$ abundance uncertainty prior to our measurement. Our experiment dramatically reduces mass uncertainties of the $N=42$ isotones between Zr and Mo in the $rp$-process, reducing the associated $A=82$ and $A=83$ abundance uncertainties at the 3$\sigma$ level to a negligible 8\% each. The final abundances calculated with our work exclude the existence of an $A=82$ abundance peak and thus the existence of a ${}^{82}{\rm Zr}$ waiting point.

In summary, we performed mass measurements of ${}^{78}$Y${}^{\rm m}$, ${}^{79}$Y, ${}^{83}$Nb, ${}^{84}$Mo, and ${}^{88}$Ru. For ${}^{78}$Y${}^{\rm m}$, ${}^{84}$Mo, and ${}^{88}$Ru, their masses are experimentally determined for the first time. The mass precisions of ${}^{79}$Y and ${}^{83}$Nb are improved to $13~{\rm keV/c^2}$ and $9.6~{\rm keV/c^2}$, respectively. The new $S_{\alpha}({}^{84}{\rm Mo})$ value proves the existence of the pronounced island of low $S_{\alpha}$ in the neutron-deficient Mo isotopes. However, the island is not strong enough for the ZrNb cycle to form in the $rp$-process. The ZrNb cycle in the X-ray bursts can, therefore, be ruled out unambiguously.  An X-ray burst simulation with the new masses shows that our measurements effectively remove the large final abundance uncertainties in the $A=80-90$ mass region caused by the previous mass uncertainties. The new mass values improve the prediction power for the composition of the nuclear ashes in X-ray bursts beyond $A=80$. \\

\textit{Acknowledgments}--We are grateful to the RIKEN Nishina Center for Accelerator-based Science and the Center for Nuclear Study at the University of Tokyo. This work was supported by the Japan Society for the Promotion of Science, KAKENHI, Grant Numbers 17H06090,  20H05648, 22H01257, 22H04946, 25H01273, 19K14750, and 21K13951, and also RIKEN programs (r-EMU and RiNA-Net). HS acknowledges support from the US National Science Foundation under Grant No. PHY-2209429.

\bibliography{rp-mass}

\end{document}